# Dimensionality-Reduction in the *Drosophila* Wing as Revealed by Landmark-Free Measurements of Phenotype

Vasyl Alba[1,2,‡], James E. Carthew[1], Richard W. Carthew[2,3], and Madhav Mani[1,2,3,*]

[1]Department of Engineering Sciences and Applied Mathematics, Northwestern University, Evanston, IL 60208

[2]NSF-Simons Center for Quantitative Biology, Northwestern University, Evanston, IL 60208

[3]Department of Molecular Biosciences, Northwestern University, Evanston, IL 60208

‡ vasyl.alba@northwestern.edu

* madhav.mani@gmail.com

## ABSTRACT

Organismal phenotypes emerge from a complex set of genotypic interactions. While technological advances in sequencing provide a quantitative description of an organism's genotype, characterization of an organism's physical phenotype lags far behind. Here, we relate genotype to the complex and multi-dimensional phenotype of an anatomical structure using the *Drosophila* wing as a model system. We develop a mathematical approach that enables a robust description of biologically salient phenotypic variation. Analysing natural phenotypic variation, and variation generated by weak perturbations in genetic and environmental conditions during development, we observe a highly constrained set of wing phenotypes. In a striking example of dimensionality reduction, the nature of varieties produced by the *Drosophila* developmental program is constrained to a single integrated mode of variation in the wing. Our strategy demonstrates the emergent simplicity manifest in the genotype-to-phenotype map in the *Drosophila* wing and may represent a general approach for interrogating a variety of genotype-phenotype relationships.

## Introduction

Discovering the properties of an organism's genotype-to-phenotype (G2P) map is fundamental to our understanding of development and evolution. As such, the physical phenotypes of an organism emerge from a complex set of genotypic interactions, precluding a straightforward path to relating the two since variation in genotype can potentiate unpredictable cellular behaviours and traits[1]. Intuiting an emergent property of an organism's G2P map,

C.H. Waddington conjectured that "developmental reactions, as they occur in organisms submitted to natural selection...are adjusted so as to bring about one definite end-result regardless of minor variations in conditions during the course of the reaction"[2,3]. Such robustness criteria suggest that only a small portion of phenotypic space is made accessible by the developmental program. In particular, the natural variation in a population, as well as small-effect genetic and environmental perturbations, are predicted to have congruent effects on adult phenotype[4,5]. If true, Waddington's conjecture would embody a striking example of dimensionality reduction in an organism's G2P map[1,6], meaning that high-dimensional genome-wide and environmental variation is mapped onto low-dimensional variation in body form through the process of development. This would provide insight into a class of general design principles and call for an investigation into its evolutionary origins and consequences[7,8].

State-of-the-art techniques to quantitatively relate genotype to phenotype are limited by current assays of physical traits. These are primarily based on an arbitrary fragmentation of a body structure into a handful of composite measurable traits, referred to as landmarks[9-11]. Based on a landmark-free mathematical approach that enables a robust description of biologically salient phenotypic variation, we develop a framework for relating genotype to the complex and multi-dimensional phenotype of a major anatomical structure. We chose the wing of *Drosophila melanogaster* as the prototype sub-system to analyse for several reasons. First, the developmental processes to form the wing are comprehensively understood, with many genes implicated in wing development[12-15]. Second, the wing has a complex form and function (Figure 1a), with a clear fitness impact for the organism[16-18]. Lastly, wing phenotype has been extensively studied using landmark-based morphometrics (Figures 1a and S1)[19-23]. Leveraging our approach, we demonstrate that the spectrum of phenotypes generated by the wing's developmental program is highly constrained, as Waddington anticipated, aligning with each other to define a single direction of congruent effects. Strikingly, phenotypic variation along this direction is not limited to isolated discrete landmarks of the wing. Instead, it makes itself manifest as an integrated mode across the entire wing, highlighting the complex and collective nature of its G2P map. Given the ease with which genomic variation can be measured, the strategy described here may represent a general approach for interrogating a variety of genotype-phenotype relationships.

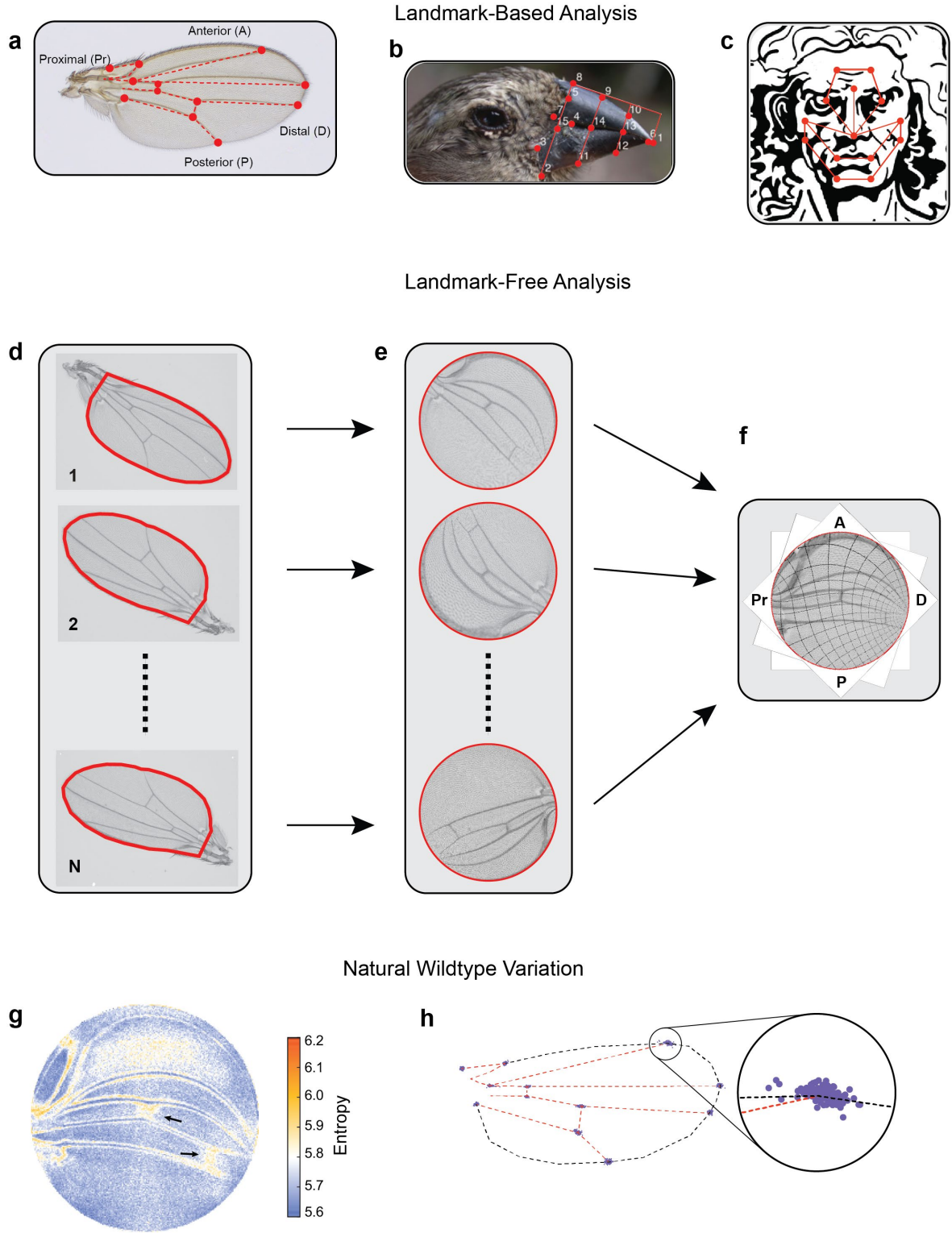

**Figure 1. Landmark-free morphometrics. a-c,** Standard landmarks (red) for Procrustes analysis of a *Drosophila* wing (**a**), the beak of a Darwin finch (**b**), and the face of DaVinci's Vitruvian Man (**c**). **d-f,** The landmark-free method involves boundary identification of wings (**d**), boundary alignment through conformal mapping to unit discs (**e**), and bulk alignment of wings through optimization in the space of conformal maps from the disc to itself (**f**). **g,** Pixel entropy of an ensemble of wings from the outbred wildtype population. Enhanced variation of cross-veins is observed (arrows) **h,** Procrustes analysis of the same ensemble of wings, highlighting variation in the 12 landmark positions. Inset, the landmark where the L1 and L2 veins intersect.

## Results

**Landmark-free morphometrics:** Current approaches to phenotyping are particularly problematic for body structures with complex form and pattern, where spatial information is discretized into landmarks - anatomical loci that are homologous in all individuals being analysed (Figure 1a- c)[24,25]. Analysis of phenotypic variation has additional challenges owing to the inability to perform precision alignment of complex two- and three-dimensional shapes, which cannot be achieved through simple schemes that account for rotation, scaling, and shear transformations [9-11].

To overcome these problems, we have developed an alignment tool that admits a landmark-free method to measure the physical totality of all traits in a body sub-system. Choosing the wing of *Drosophila melanogaster* as the prototype sub-system, the landmark-free method is comprised of four steps (Figure 1d-f). First, each wing is imaged by transmitted light microscopy at high resolution (Figure S3b). Second, the boundary of the wing image is accurately detected (Figure S2)[26]. Third, the boundary of the wing image is computationally mapped to the interior of a fixed-sized disc. The mapping of the boundary to the disc relies on an efficient numerical implementation of the Riemann Mapping Theorem[27-30]. In particular, the map is conformal, preserving the shape of the image, locally, by preserving angles, while manifestly distorting areas (Figure S4). This feature of the approach is both a strength and a limitation, down-weighting the effects of variation in boundary shape, and up-weighting the effects of the pattern of hairs and veins in the interior of the wing (Figure S4j,k). Fourth, mapped images are globally registered to one another (Figures 1e-f and S4l), since disc-mapped wings vary in their orientation and region of focus (the central point in the disc). These misalignments are spanned by the space of conformal maps of the disc onto itself, which can be optimized to produce a global registration of the images, as shown in Figure 1f (further details can be found in Methods). We can thus generate boundary and bulk registered ensembles of wing images, whose variation we can study at single-pixel resolution in a landmark-free manner.

We imaged wings from a highly outbred population of wildtype *Drosophila*. This population was founded from 35 inbred strains collected worldwide, which were then blended together for two years to ensure substantial genetic variation. We visualized the quantitative variation in the ensemble of images at the single-pixel resolution, measured as per-pixel information entropy (Figure 1g)[31]. Entropy is a measure of the uncertainty, which here is based on variations in image intensity determined on a per-pixel basis[32]. Regions of higher entropy straddle the longitudinal veins, indicating that the positions of these veins along the anterior-posterior (AP) axis vary slightly between individuals; less than a single vein width across the ensemble of wings. However, the proximal region of the L1 vein shows considerable variation, as do some other regions near the hinge. In addition, the cross-veins substantially vary between individuals along the proximal-distal (PD) axis. The intervein region between L1 and L2 veins and the intervein regions near the distal tip also show variation owing to variation in wing hair location. A Procrustes analysis of the same wing ensemble is limited to the variation in landmarks alone, with the intersection of L1 and L2 veins being the most variable (Figure 1h and S1c). An anatomical description of the *Drosophila* wing can be found in Figure S3c-d.

The wings of male and female *Drosophila* differ in their size, shape, and pattern (Figure S5a-b), providing a setting to compare the landmark-based and landmark-free approaches to distinguishing two populations. By Procrustes landmark analysis, the intersection of L1 and L2 veins shows the greatest sexual dimorphism, with substantial non-overlap between male and female (Figure 2a). The intersection of L1 and L5 veins also shows sexual dimorphism. A very different perspective is revealed using our landmark-free method (Figure 2b). The relative entropy of pixel intensities across the two ensembles demonstrates that sexual dimorphism is not restricted to the landmarks. In particular, the proximal and distal regions of the L1 vein are variable between male and female. The L3 vein proximal to the cross-vein also shows considerable variation. Strikingly, the regions showing sexual dimorphism do not spatially overlap with the regions showing variation within each wildtype population. In summary, this comparative analysis demonstrates the remarkable sensitivity of our landmark-free method to make precise measurements of the variational properties of wing form, as well as demonstrating the incomplete picture of variation provided by landmarks alone.

**A Dominant Mode of Natural Phenotypic Variation:** We note that the space in which phenotypes are parameterized via our landmark-free approach is vast. The intensity of each pixel in the space of disc-mapped images is identified with a unique dimension (Figure 2c). Therefore, an individual image is a single point in this phenotype space of ~30,000 dimensions (pixels). By contrast, the standard landmark-based parameterization of the wing defines phenotype in 24 dimensions - the *x* and *y* locations of 12 landmarks. By way of illustration, we portray an ensemble of wildtype wings populating a three-dimensional slice of this vast phenotypic space (Figure 2c). The dominant mode of variation in the population is indicated as the "long-axis" of the cloud of points. To find the major axes of variation within the 30,000-dimensional space, we performed principal components analysis (PCA) on male and female wings from the outbred stock raised under standard environmental conditions. PC1 and PC2 axes account for 68% of the total variation, with 66% of the variation aligned along PC1 (Figure 2d). Strikingly, male and female wings are separated along PC2, suggesting that the dominant mode of variability is not sex-specific. This dominant mode of variability is also not congruent with left-right patterning (Figure S5d), variation in wing size (Figure S5e), or variation in imaging conditions across the ensemble (Figure S5f). We mapped pixels whose variation is most aligned with PC1 and found vein positions are most prominent, particularly in the proximal region (Figure 2e). We also performed PCA on the Procrustes landmark data from the wildtype ensemble and observed that sexual dimorphism is the dominant source of detected variation (Figure 2f). Strikingly, the dominant mode made manifest in our landmark-free analysis is undetected by a landmark-based method (Figure S5c), demonstrating that novel, even dominant, modes of variation are rendered detectable by our more holistic analysis.

**Genetic and Environmental Variation:** To investigate the origins of this dominant mode of variation, we performed landmark-free analysis on a published dataset of wing images from a wildtype Samarkand strain and from animals heterozygous for weak mutations in signal transduction genes[33]. Specifically, these genes act in the Notch, BMP, and EGF Receptor (EGFR) pathways, which are required for proper growth and patterning of the developing *Drosophila* wing[15]. Note that all of the mutations are considered to be recessive to wildtype for wing phenotype, and thus the heterozygous animals do not exhibit easily discernible differences in wing form from wildtype. Since stronger effect mutations produce phenotypes rarely seen in a natural population, the weak nature of these genetic perturbations is crucial to our investigations into the origins of natural variation.

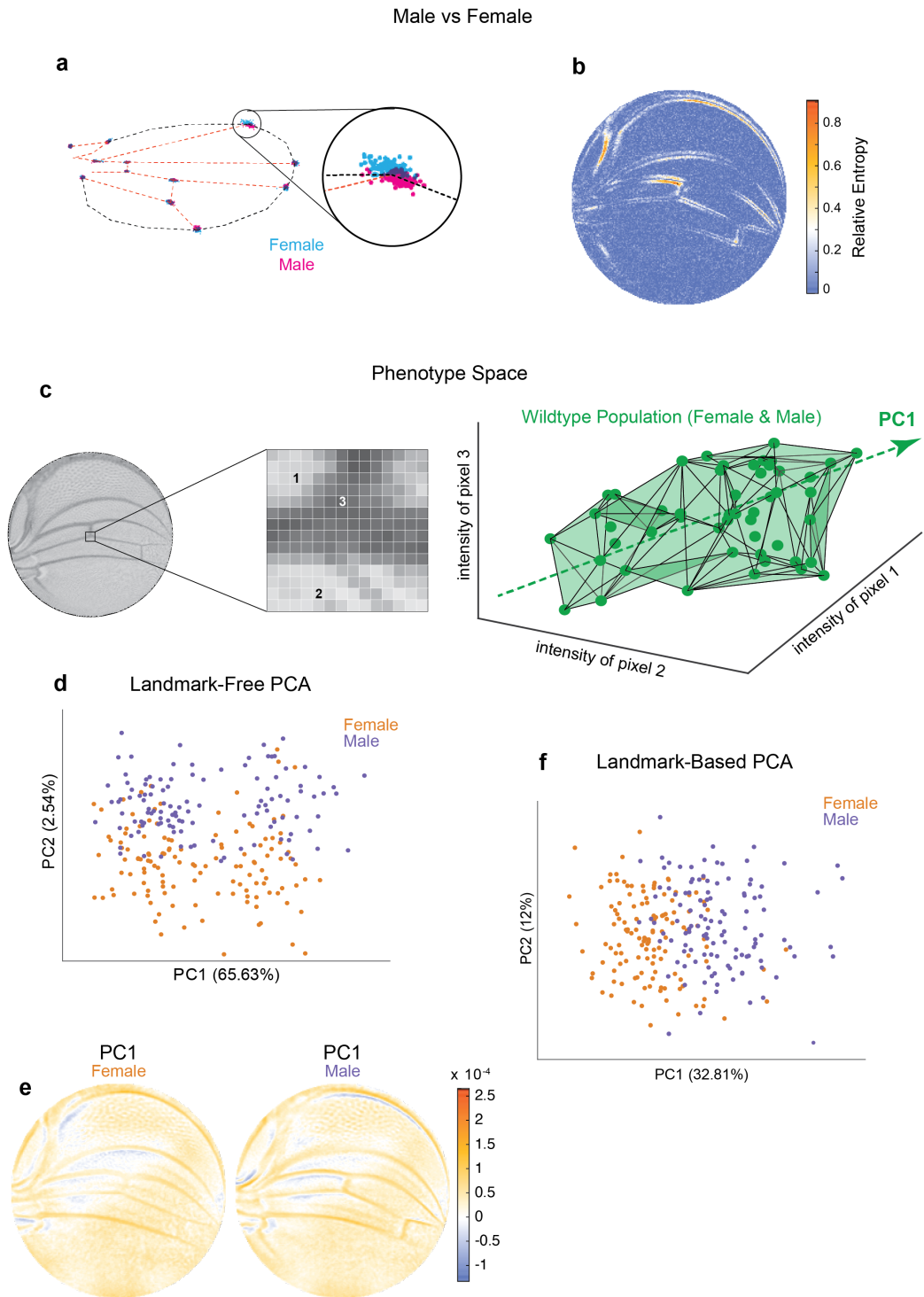

**Figure 2. Comparison of Procrustes and landmark-free phenotyping. a-b,** Variation between ensembles of male and female wings from the outbred wildtype population. (**a**) Procrustes analysis with inset showing the landmark where L1 and L2 veins intersect. (**b**) Per-pixel symmetrized Kullback-Leibler divergence in landmark-free analysis detects variation along veins, undetected by landmark-based analyses. **c,** The dimensions of phenotype space comprise individual pixels. Three such pixels are randomly selected here to show their 3D phenotype space. An ensemble of wings are points in this space. The direction of largest variation in the space is identified by PCA as PC1. **d,** Landmark-free PCA analysis of the outbred wildtype population reveals a novel dominant direction of variation orthogonal to sex-specific variation. Sex-specific variation aligns with PC2. **e,** Pixel alignment with PC1 analysis shown in panel d, showing vein position variation aligns with PC1. **f,** Procrustes-based PCA analysis of the same ensemble analysed in panel d. PC1 only detects the sex-specific variation.

Mutant wing images were subjected to our landmark-free analysis. The high-dimensional phenotype space occupied by each mutant population was compared to the reference wildtype population. This was done by finding the centroid (centre of mass) for each population's cloud of points, which constitutes its average phenotype. We then considered the line joining each mutant centroid and the reference wildtype centroid, which represents the difference between the average mutant and wildtype phenotypes. For illustration, a simplified three-dimensional example is shown in Figure 3a. The spatial maps of the mutant-wildtype differences in pixel intensities are revealing (Figure 3b-e), and they show how our method can detect even subtle effects (see Methods). Mutation of the *Egfr* gene causes L3 and L4 veins to be slightly more distant from one another, whereas mutation of *Star*, which also acts in the EGFR pathway, has the opposite effect. The posterior cross-vein is also affected by these mutants. Both *mastermind (mam)*, a component in the Notch pathway, and *thick veins (tkv)*, a component in the BMP pathway, have common effects in that L2 - L4 slightly shift anterior. However, *tkv* shows evidence of subtle vein thickening, which is an attribute of the homozygous mutant phenotype. We conclude that the landmark-free method is sensitive enough to robustly detect phenotypic differences in mutant alleles that are historically considered to be completely recessive to wildtype.

How does the phenotypic variation in the mutants compare with the phenotypic variation in the outbred population? We measured the alignment between the top four vectors of variation in the outbred population (PC1-PC4), and the vectors connecting Samarkand and mutant centroids (Figure 3a). The cosine of the angle $\theta$ between each PC vector and each centroid-centroid vector is reported in Figure 3f. There is a high degree of alignment between the dominant mode of variation (PC1) in the outbred population and almost all of the mutants ( $\cos(\theta) \rightarrow 1$ as $\theta \rightarrow 0$). With the exception of the male population of *Star* heterozygotes, the degrees of alignment are far above those expected by chance due to sample size effects or artefacts in the imaging and analysis pipeline (Figure 3g). As such, the mutant phenotypes statistically align with a single direction, which is identified as the direction of dominant variation in an outbred wildtype population. We note that the mutants affect distinct signalling pathways, which play significant and separate roles in wing development, thus highlighting the nontrivial nature of the statistical alignment we observe. Taken together, these small-effect genetic perturbations – both the recessive mutations and the variants in the outbred populations – have a congruent effect on phenotype.

As suggested by Waddington, other types of perturbation might have similar effects. To test this, we raised the outbred wildtype population at different temperatures from the standard 25°C. As previously observed, increasing temperature causes a decrease in wing size (Figure S5g,h). The differences between the average wing phenotypes at different temperatures were spatially mapped (Figure 4a-b). We also raised the outbred population on nutritionally limited food, which decreases wing size (Figure S5i,j)[34]. The average wing phenotype due to nutrient limitation also reveals robust spatial variation (Figure 4c). Remarkably, there is a strong statistical alignment of the temperature and dietary phenotypes with the dominant mode of variation within the outbred population raised under standard environmental conditions (Figure 4d-e). Thus, small-effect environmental perturbations are also constrained in their impact on wing phenotype, aligning with the previously identified direction of congruent effects.

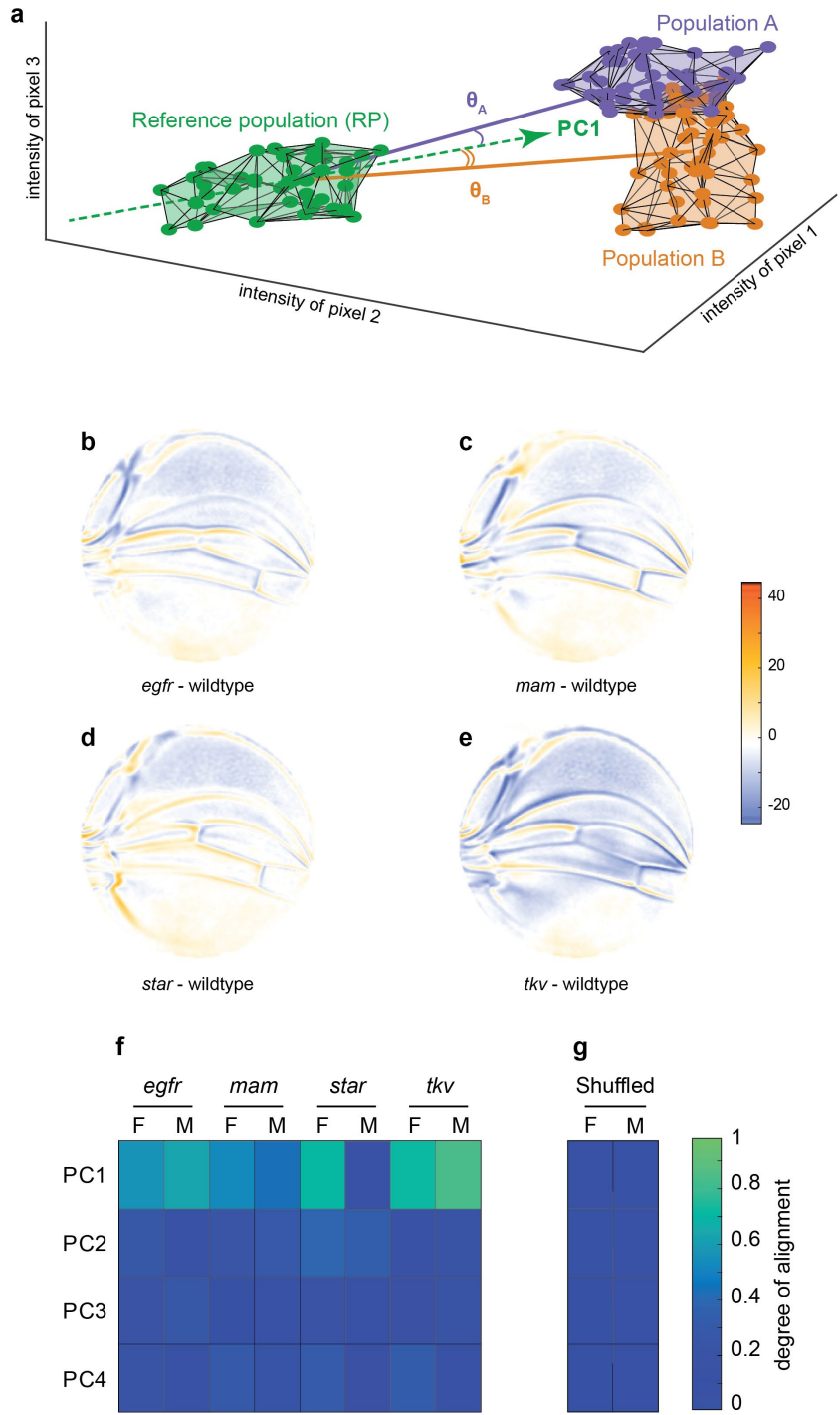

**Figure 3. Variational analysis of genetic mutants. a,** Schematic of hypothetical 3D phenotype space, with each cloud of points representing wings from an ensemble subject to a different treatment or condition. The PC1 vector for each cloud of points can be independently calculated. The vectors joining the centroids of any two clouds of points can also be calculated. The angles between the direction of maximal variation (PC1) of the reference population and the directions from the reference to test ensembles are measured ($\theta_A$ and $\theta_B$). **b-e,** Per pixel intensity difference between each mutant and the wildtype reference as measured at the centre of each point cloud. Negative values are when mutant pixel intensity is smaller than wildtype, and positive values are when mutant pixel intensity is larger. Comparison between wildtype and *egfr* (**b**), *mam* (**c**), *Star* (**d**), and *tkv* (**e**) ensembles. **f,** The cosine of the angle between the directions of PC1 – PC4 of the outbred wildtype population, and the directions of the vectors connecting the wildtype and mutant populations. M and F refer to male and female groups, respectively. **g,** The cosine of the angle when the wildtype and mutant populations analysed in panel f are shuffled but the sexes are not. Shuffling leads to little or no alignment of mutant vectors with any PC

**Geometric Analysis of Phenotypic Variation:** Why are some perturbation phenotypes, for example, *Star* mutant male wings, less aligned to the primary mode of natural phenotypic variation than others (Figures 3f and 4d)? Either phenotype is less constrained for some perturbations or some underlying feature of the data was not being considered. Here, we demonstrate that a geometric feature of the data alone explains the observed variations in alignment. In particular, variation in the distance in phenotype space separating two wing populations accounts for the variation in the observed degree of alignment.

We consider a situation where wing phenotypes of a population lie a small distance, $\sigma$, away from the axis of primary variation (PC1) of a reference group (Figure 5a). We define $R$ to be the centroid-centroid distance between the two groups, the strength of the phenotypic difference, and $\theta$ to be the angle between the axis of primary variation of the reference group and the vector connecting the centroids of the two groups, a measure of alignment. The three quantities are related through a simple trigonometric relation, $\sigma/R = \sqrt{1 - cos^2\theta}$, which mathematically embodies the intuitive fact that if small-effect perturbations can only access a small region of phenotype space perpendicular to the axis of primary variation, then the measure of alignment must increase with the strength of the perturbation. This would mean that $\sigma/R \to 0$ as $R$ increases (Figure 5a). Strikingly, this relation is observed for all of the environmental and mutant perturbations, analysing bootstrapped subsamples of the data (Figure 5b-c). The results suggest that for all of the genetic and environmental perturbations tested, they generate minor phenotypic deviation orthogonal to the primary mode of natural phenotypic variation. Most variation manifested by the perturbations align with the primary mode of natural variation. For example, the departure of *Star* mutant male wings from alignment with the primary mode of variation (Figure 3f) can be explained by the weak phenotypic difference they have, as demonstrated by the short distance between *Star* and wildtype centroids (Figure 5b). In summary, the phenotypic variation caused by weak genetic and environmental perturbations is constrained along a single linear manifold that aligns with the phenotypic variation seen within every population. Together, this indicates the presence of powerful dimensionality-reduction in the developmental program of the *Drosophila* wing.

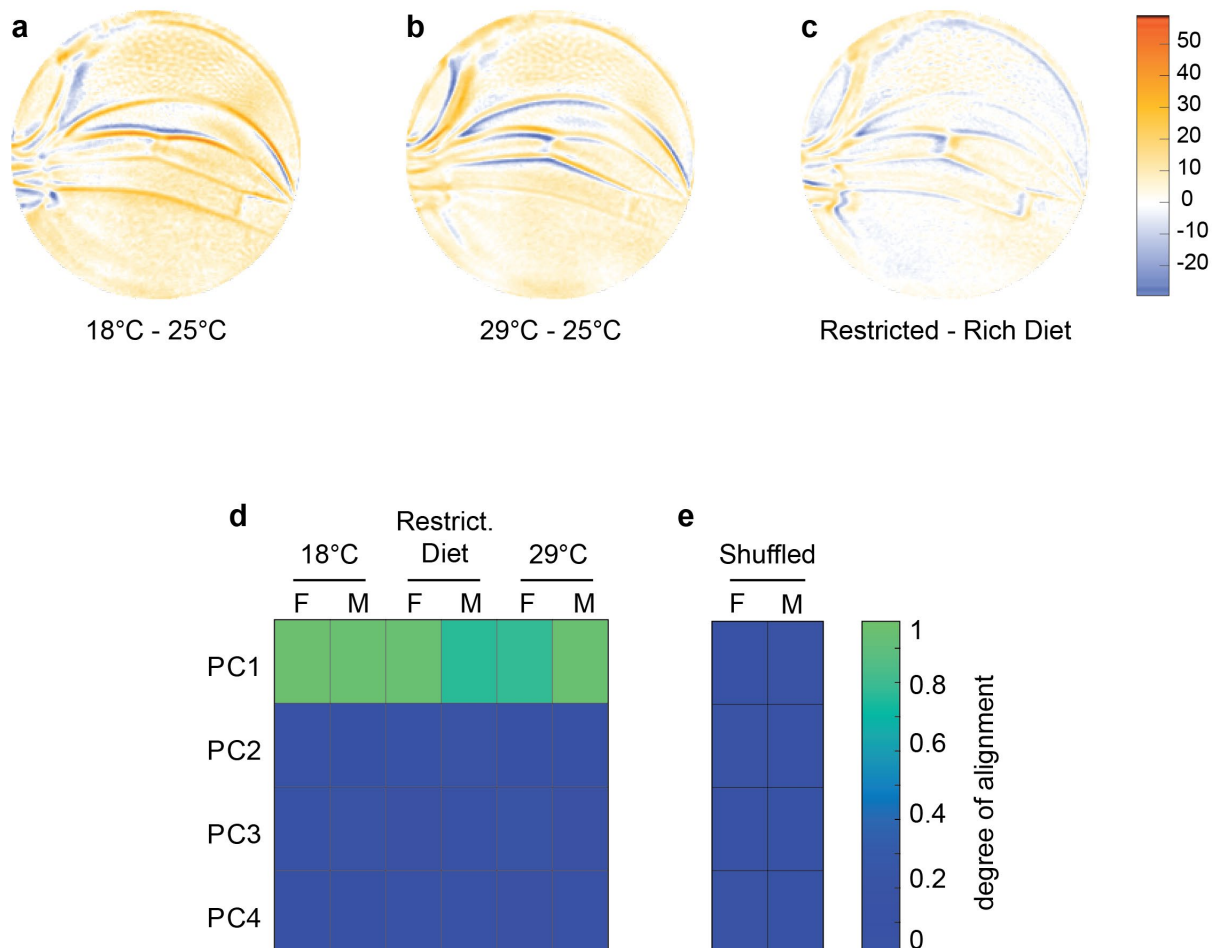

**Figure 4. Variational analysis of environmental perturbations**. **a-c,** Per pixel intensity difference between wings from a test condition (described on left) and a reference condition (described on the right). Difference is measured at the centre of each point cloud. Negative values are when test pixel intensity is smaller than reference, and positive values are when test pixel intensity is larger. Comparison between wings from 18 and 25°C (**a**), 29 and 25°C (**b**), and restricted vs rich diet (**c**) treatment. **d,** The cosine of the angle between the directions of PC1 – PC4 of the outbred wildtype population, and the directions of the vectors connecting the populations raised under different temperature and diet conditions. M and F refer to male and female groups, respectively. **e,** The cosine of the angle when the treated populations analysed in panel d are shuffled but the sexes are not. Shuffling leads to little or no alignment of vectors with any PC.

## Discussion

We observe that phenotypic variation in the *Drosophila* wing, generated by genetic and environmental variation, populates a single one-dimensional structure in phenotype space, evidence of the strong dimensionality reducing properties encoded in its developmental program. This direction emerges as an integrated and spatially-extended feature across the wing that is not encoded in the spatial fluctuations of a small number of human-specified landmarks. This single mode of variation primarily affects the position of wing veins along the AP or PD axes, particularly the position of the L1 – L4 veins close to the hinge. Final vein position can be affected by two possible developmental processes. One, thick proveins are refined into narrow veins during pupal development, allowing for some wiggle as to where the veins reside[15]. Two, the number, size and shapes of intervein epidermal cells might vary, and after expansion, separate neighbouring veins by different distances[35-38]. These effects could be exacerbated in the proximal wing during hinge contraction[39-41]. Notably, vein thickness and shape are not variable within the manifold of phenotypic variation.

We postulate that it is the continuous, landmark-free, representation of phenotype that permitted the detection of this novel structure. Our landmark-free method can be applied to a two-dimensional image of any sub-system of a body, no matter how complex. We chose the *Drosophila* wing because its flattened nature makes two-dimensional imaging quite straightforward. However, our method will work for any closed body sub-system subjected to 3D surface imaging via laser scanning or computer tomography. This will allow quantitative landmark-free phenotyping of skeletal structures, teeth, and soft organs such as the brain in living and dead specimens.

More generally, our results highlight the complex and integrated nature of the G2P map in the *Drosophila* wing. While its origins remain unclear, the compactification of the accessible portions of phenotype space, generated through the dimensionality-reducing properties of development, generate a near-continuous spectrum of phenotypes, which nature can select from and evolve.

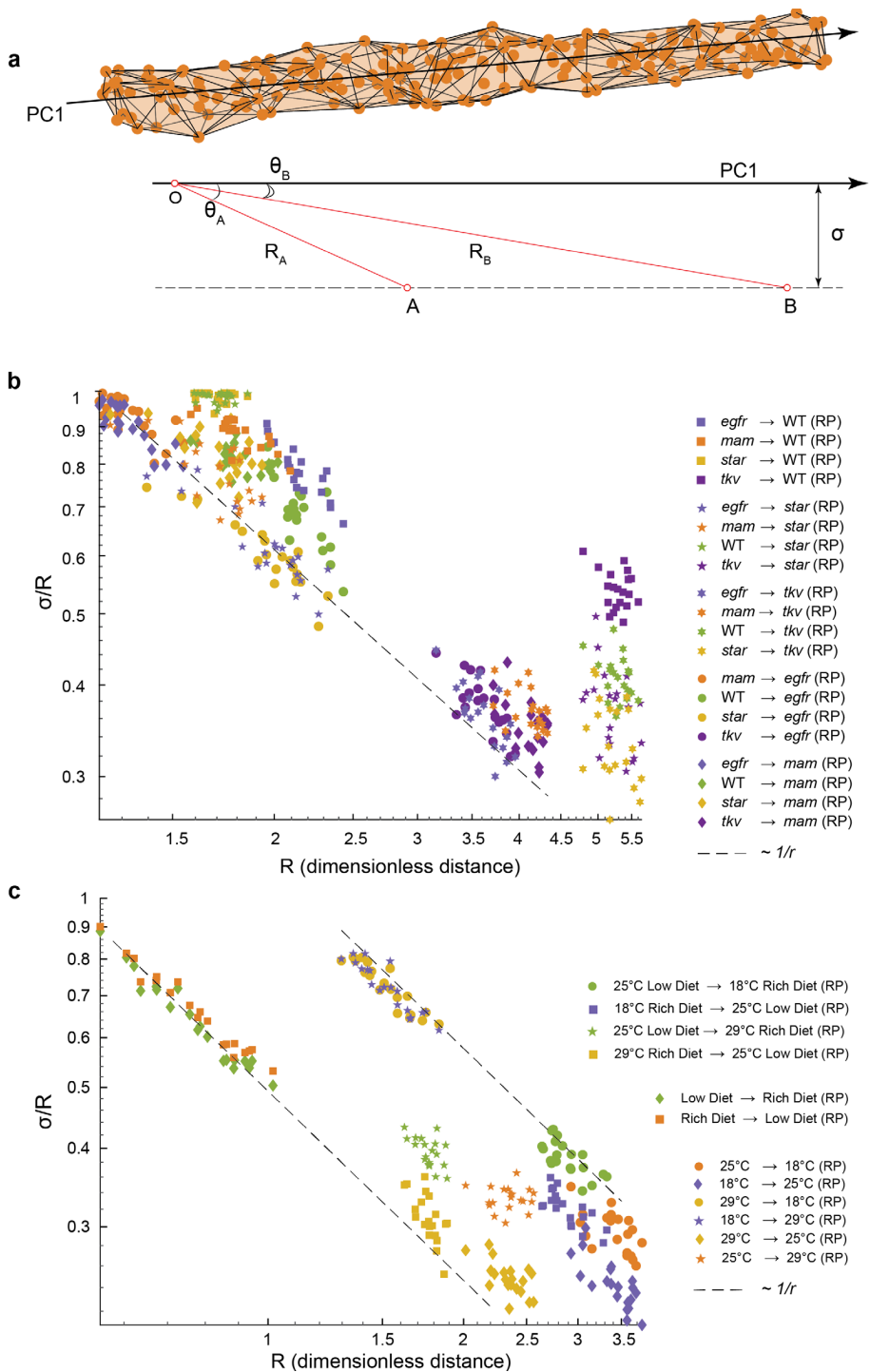

**Figure 5. Geometric analysis of dominant variational mode. a,** Schematic of a data manifold with a single dominant linear direction of variation, embedded in high-dimensional phenotype space. Variation orthogonal to the dominant direction is visualized with a characteristic scale, σ. In the presence of such a data manifold, the angle between the dominant direction of a reference population, PC1, and the centroid-to-centroid vector connecting any two populations is predicted to decrease as a function of R, the length of the centroid-to-centroid vector. **b-c,** The angle between the direction of maximal variation (PC1) in a reference population (RP) and the centroid-to-centroid vector connecting the RP and a second population as indicated. This is shown as a function of R, the distance between them. Samples are bootstrapped to indicate the in-sample variance, with each bootstrap represented by a point. Both in the mutant/wildtype (WT) (**b**) and environmental (**c**) ensembles, alignment grows as a function of distance, indicative of a single dominant direction in the data manifold. The dotted lines represent the relationship between σ/R versus R if sigma was a constant.

**Acknowledgements:**

We thank Seppe Kuehn for fruitful conversations. This work was supported in part by NSF grant DMS-1547394 (JEC), NSF (1764421, MM and RWC), and the Simons Foundation (597491, MM and RWC). MM is a Simons Foundation Investigator.

**Author contributions:**

Vasyl Alba: Conceptualization, Experimental design, Methodology, Data acquisition, Analysis, Writing
James Carthew: Experimental design, Data acquisition, Analysis
Richard Carthew: Conceptualization, Supervision, Funding acquisition, Methodology, Writing, Project administration
Madhav Mani: Conceptualization, Supervision, Funding acquisition, Methodology, Writing, Project administration

**Competing Interest Statement**

The authors have declared no competing interest.

# Methods for "Dimensionality-Reduction in the *Drosophila* Wing as Revealed by Landmark-Free Measurements of Phenotype"

Vasyl Alba[1,2,‡], James E. Carthew[1], Richard W. Carthew[2,3], and Madhav Mani[1,2,3,*]

[1]Department of Engineering Sciences and Applied Mathematics, Northwestern University, Evanston, IL 60208

[2]NSF-Simons Center for Quantitative Biology, Northwestern University, Evanston, IL 60208

[3]Department of Molecular Biosciences, Northwestern University, Evanston, IL 60208

‡ vasyl.alba@northwestern.edu

* madhav.mani@gmail.com

**Fly Husbandry**

Cornmeal-molasses food was used for standard husbandry. Inbred lines were kept in bottles or vials. The outbred population was kept at 25°C in an 8.8 litre population cage. The size of the cage and food availability limited the population size to 1,000 − 2,000 adults per generation. Populations were raised in Percival incubators. The temperature of incubators was monitored daily. Fly bottles or vials were monitored for moisture levels. Any condensed moisture on the inside of the container walls was absorbed using Kimwipes. Incubators were monitored for mite infestations using mite traps checked daily.

**Construction of the Outbred Wildtype Population**

A collection of 35 wildtype *Drosophila melanogaster* stocks were obtained from the Bloomington Drosophila Stock Center (BDSC). These inbred stocks originated from founder individuals isolated from around the world (Excel File 1). Fourteen stocks were established in the 20[th] century from four continents plus the island of Bermuda: 1 from Africa, 3 from Asia, 1 from Bermuda, 4 from Europe and 5 from North America. The other 21 stocks were established in 2003 as isogenic lines from the Drosophila Genetics Resource Panel (DGRP). Founders of the lines were isolated in the area of Raleigh, North Carolina, USA.

Six of the DGRP isogenic stocks were randomly paired with one another, and 20 males from one stock were mated with 20 females from the other stock (Excel file 2). The offspring from these 3 crosses were then used in addition to the remaining 29 inbred stocks for a new round of pairwise mating - 16 F0 crosses in total (Excel file 2). The assignment of stocks to pair was made randomly, as were the assigned sexes used in each paired mating. Again, 20 males from one were mated with 20 females from the other. The offspring from these 16 crosses were used for another round of pairwise mating - 8 F1 crosses in total (Excel file 2). The assignment of stocks to pair was made randomly, as were the sexes used in each mating. For these crosses, 50 males from one were mated with 50 females from the other. The offspring from these 8 crosses were used for an F2 round of pairwise mating - 4 crosses in total (Excel file 2). The assignment of stocks to pair was made randomly, as were the sexes used in each mating. For these crosses, 100 males from one were mated with 100 females from the other. The offspring from these 4 crosses were used for an F3 round of pairwise mating - 2 crosses in total (Excel file 2). The assignment of stocks to pair was made randomly, as were the sexes used in each mating. For these crosses, 300 males from one were mated with 300 females from the other. The offspring from these 2 crosses were used for a final round of mating. For this cross, 800 males from one were mated with 800 females from the other. Mating was performed in a population cage. The population cage was 8.8 litres in volume. Offspring from this final cross constituted the outbred stock, the net result of the systematic mixing of 35 distinct populations. The outbred stock was then propagated in population cages for another 50 generations. The purpose of this was for the population to approach equilibrium in allele frequencies.

### Temperature and Diet Perturbations

Twenty-five mating pairs of outbred adults were mated for two days in a bottle, and offspring were raised at a uniform temperature of $18°C$, $25°C$, or $29°C$ in Percival incubators. Standard cornmeal-molasses food was used. To compare the effects of rich versus limited diet, 25 mating pairs of the outbred population were mated for two days in a bottle, and offspring were raised on food with low yeast content. We used the food recipe described in[34]. The low-yeast food recipe is: 1 L food contains 20 g dried bakers' yeast (Bakemark), 100 g sucrose (Sigma), 27 g Bacto Agar (Difco), 3 mL propionic acid (Sigma), and 30 mL 10%(w/v) Tegosept in ethanol (Sigma).

### Wing Mounting

Young adults were sorted by sex and stored in 70% (v/v) ethanol for a minimum of 24 hours. Flies were discarded if they had a torn or wrinkled wing. A single fly was dissected in a few drops of 70% ethanol solution on a microscope slide. Wings were removed by clasping the wing joint with a pair of forceps and pulling the wing from the body, using a different pair of forceps to hold the body in place. Both left and right wings were dissected, and the carcass was discarded. Both wings were then placed in a small volume of 70% ethanol solution on the microscope slide ventral side down. A few drops of 70% (v/v) glycerol were added to the wings/ethanol on the slide. If the alula was folded onto the dorsal side of a wing it was unfolded or removed. A coverslip was placed over the two wings, and any bubbles under the wings were removed by gently pressing down on the coverslip with a pair of blunt forceps. Only the left and right wings from one adult were mounted per slide.

### Microscopic Imaging

Imaging was performed using a Zeiss Axioplan microscope. A 4x 0.1NA Plan Apochromatic objective was used and the Optivar was set to 1.25x magnification. The camera adaptor magnification was 0.5x. This gave a total magnification of 2.5x. The condenser was focused and centred on a wing specimen, and then the stage was moved in order to define shading correction at a neutral point with no slide or wing in the image. Specific shading correction was defined before any images were taken in an imaging session. Each wing had two images taken;

one with unfiltered light from the microscope's halogen lamp at 100% power, and one with the green filter in place between the lamp and specimen. The slide remained in the exact same position for each of the two images. After one wing was imaged, its partner wing was then imaged. Wing images were taken so that the anterior side of the wing faced downwards in the image so as to eliminate confusion between right and left wing labelling. If either wing appeared damaged or if dust or debris marked either wing, we rejected that wing and its partner wing. For some female wings, the specimen was too large to be captured in one image. For these, we captured a portion of each wing in one image and then moved the stage to capture the rest of the wing in a separate image. The two images had sufficient overlap that they could be computationally stitched together post-imaging.

Images were captured as CZI files with a Zeiss Axiocam ERc 5s attached to a Dell XPS 8500 computer using ZEN Blue 3.1 imaging software. Each image had dimensions of 2560 x 1920 pixels, and each pixel intensity was recorded in 8-bit RGB mode. Each CZI file was converted to a TIFF file using the Bio-Formats 6.5.1 plugin for Matlab. Exposure time was 0.3 ms for white-lit images and 0.4 ms for green-lit images. Colour balance was set to auto. The intensity level was set at 100%. This ensured that the background was not saturated, i.e. background pixel intensity was not maximal. The average background pixel intensity was reproducibly between 210 and 215 for almost all wing samples (Figure S5e). For those wing samples that were captured using two overlapping images, we used the Photomerge algorithm in Adobe Photoshop CS3 with default settings. Although both white-lit and green-lit images were captured for each wing, we only used the green channel of the green-lit RGB files for further analysis.

We analysed both left and right wings for each individual. The datasets generated for the following conditions were: 18°C (169 females and 71 males), 25°C (202 females and 202 males), 29°C (185 females and 208 males), 25°C limited diet (216 females and 219 males).

**Image Analysis of Genetic Mutants**

We used a publicly available image database of *Drosophila melanogaster* wings that had been used for landmark analysis by others [33]. Briefly, P-element insertion mutants in the *Epidermal growth factor receptor (Egfr), mastermind (mam), Star (S), and thick veins (tkv)* genes had been extensively backcrossed into the Samarkand wildtype strain in order to introgress their genetic backgrounds with Samarkand. Each P-element mutant is described as hypomorphic when homozygous, meaning they are weak loss-of-function mutations. They were crossed to the Samarkand strain to generate heterozygous mutant offspring. These were the wings that had been imaged in addition to fully wildtype Samarkand wings, and all were deposited in a public database[42].

We rejected some images due to quality issues that included but were not limited to dust, non-homogeneous illumination, damage to the wings. The remaining images were used for landmark-free analysis: Wildtype (214 females and 200 males), *Egfr* (231 females and 236 males), *mam* (212 females and 260 males), *S* (230 females and 222 males), and *tkv* (232 females and 232 males).

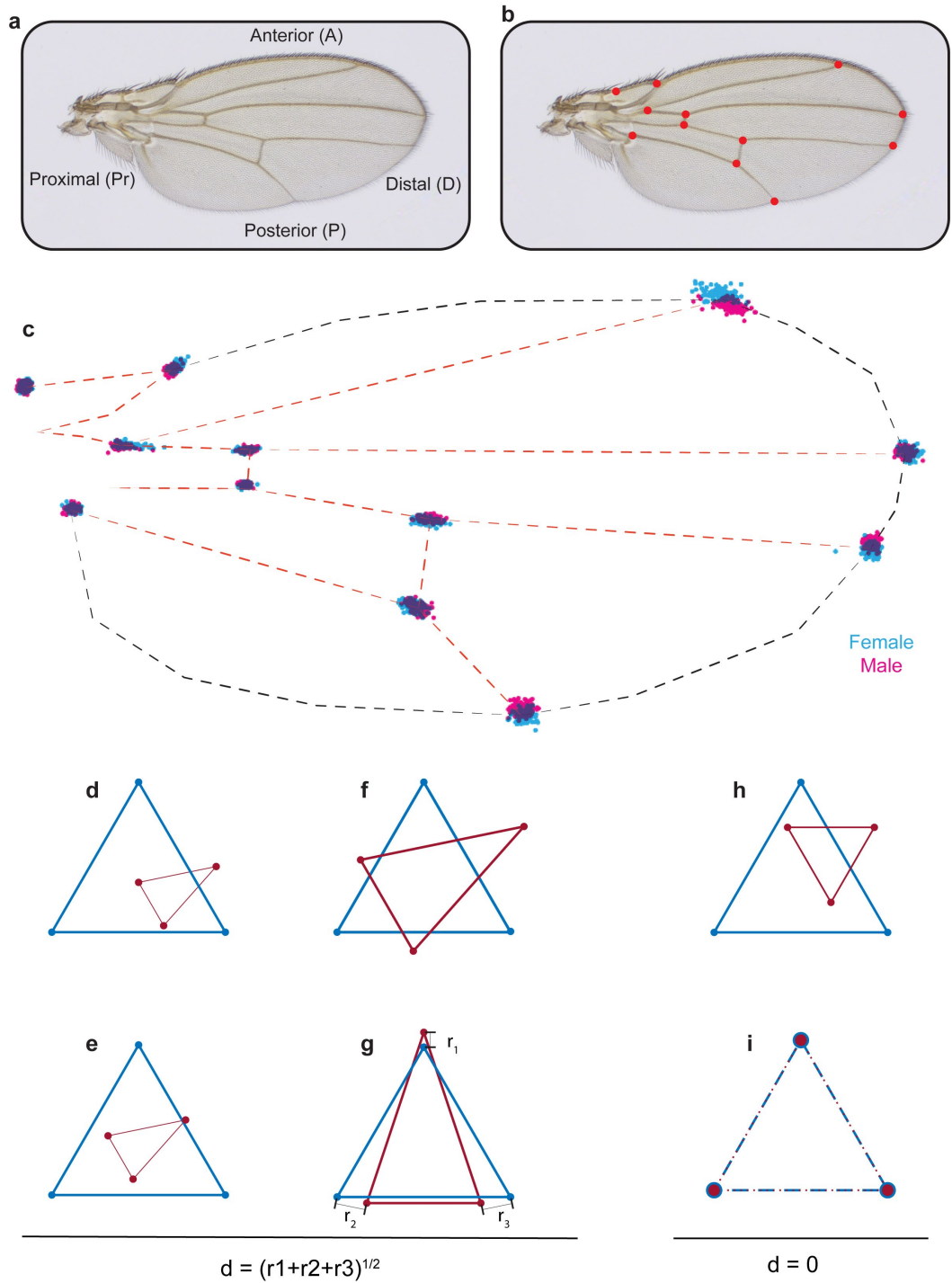

$$d = (r1+r2+r3)^{1/2} \qquad\qquad d = 0$$

**Figure S1. Procrustes alignment of landmarks. a,** Image of a *Drosophila* wing with axes labelled. **b,** Image of a wing with the 12 standard landmarks used in Procrustes analysis. **c,** Positions of landmarks of outbred wildtype male and female wings after Procrustes alignment. **d-g,** In order to explain Procrustes alignment, we use a simple example with two triangles of different size, shape and orientation (**d**). **e,** The first step is to align their centroids. **f,** The second step is to scale the two shapes to make them of equal area. **g,** The last step is rotation to optimally align the shapes. **h-i,** The simplest case is when the triangles have identical shape (**h**), and as a result, Procrustes alignment generates complete overlap of the vertex landmarks (**i**).

**Landmark-Based Morphometrics**

Landmarks are defined as anatomical loci that are unambiguously identifiable and are homologous in all individuals being analysed. In the case of the *Drosophila melanogaster* wing, 12 points where wing veins intersect have been historically used as landmarks (Figure S1a). We used the Wings 4 software tool that had been developed specifically to automatically identify the position of the 12 landmarks on each wing. The software fits a spline model of the wing to each picture of the wing. The software and manual can be found at[43].

Once the landmarks were identified for each image, the ensemble of landmarks was aligned using a standard Procrustes method[44]. In the Procrustes analysis, objects are superimposed by optimal translation, rotation and uniform scaling of the objects. Both the placement in space and the size of the objects are freely adjusted. The aim is to obtain a similar placement and size by minimising the Procrustes distance between the objects.

As a hypothetical example, we take a sample triangle and align it with a reference triangle (Figure S1d-g). Once the shapes are on top of one another, the sample shape is uniformly scaled to match the area of the reference shape. The last step is to rotate the sample shape to best-align with the reference. These steps are performed until the average distance between the landmarks (vertices of the triangle) of the sample are minimally distant to the landmarks of the reference.

A measure of the goodness of the superimposition is the Procrustes distance:

$$d = \sqrt{(x - x_{ref})^2 + (y - y_{ref})^2 + \cdots},$$

where (*x, y, . . .*) and (*x*$_{ref}$, *y*$_{ref}$, . . . ) are coordinates of the landmarks of the sample image and reference image, respectively. There are 12 landmarks on the *Drosophila* wing, thus there are 24 x and y coordinates that represent each wing. In the case of triangles, $d = \sqrt{r_A^2 + r_B^2 + r_C^2}$ (Figure S1d-g). If two shapes are exactly the same but have different orientation and position, the Procrustes method will perfectly align them (*d* = 0) (Figure S1h-i). For the *Drosophila* wing landmarks, we used the Wings 4 software to perform the Procrustes alignment of the landmarks.

Since there are 24 *x* and *y* coordinates that represent each wing, one can think about each wing as a vector in 24−dimensional space. One can perform Principal Components Analysis (PCA) on these vectors for an ensemble of wings. As a result, one can get directions of the largest variation in landmark positions, but there will not be any information about the positions of the veins or other parts of the wing that have no landmarks. We performed PCA on the landmark-based data using the `pca` function from the Dimensionality Reduction and Feature Extraction toolbox in Matlab.

**Landmark-Free Morphometrics**

The main concept of the method is to treat every pixel in a wing image as a variable, a trait, with a discrete intensity value. Hence, the higher the resolution of the image, the greater the number of variables. The resolution of wing images we captured are such that there are 100,000 pixels for each wing. However, the pixels are not landmarks in the traditional sense. Traditional landmark-based methods globally align identifiable anatomical loci from homologous objects of different individuals. Then, the variation in the landmark coordinate position is measured for each landmark. Our landmark-free method globally aligns the entire image from homologous objects of different individuals. Then, the variation in intensity for each pixel coordinate is measured.

Homologous body structures from different individuals can have different shapes. This can apply not only to individuals from different species (i.e. forelimb of bat and horse) but also individuals from the same species. This is why the field of morphometrics has reduced the complexity of the continuum of shape to a discretized proxy, i.e. landmark-based Procrustes analysis then can estimate shape variation.

Our approach is to use a conformal map to map all homologous body structures onto the same shape, in this case, a disc of fixed size. Although we apply the approach to one body structure, the *Drosophila* wing, in principle it can be applied to any shape. This could be a simple 2D shape like the wing or the closed surface of a complex 3D shape such as a tooth, brain, or another body part.

**Wing Boundary Identification**

Although the background intensity of all images was highly consistent across ensembles (Figure S5e), there were minor differences, including uneven illumination across the field of view. This variability was most problematic with the wing images from the genetic database[42]. To correct these issues we apply contrast-limited adaptive histogram equalization (CLAHE) to our images (Figure S3a)[45].

We use an ilastik segmentation model to identify pixels that belong to the wing bulk, veins and background (Figure S2a). The machine learning model is trained on 25 wings and after that is applied to all wings across all ensembles to remove systematic trends owing to variations in segmentation. The output of Ilastik comprises four layers (Figure S2b). The segmentation output requires further cleaning that is done in a custom-made Matlab script. In particular, we use knowledge about the fact that background is only outside of the wing, and wing bulk is surrounded by veins, to clean up the segmentation (Figure S2c). Once the segmentation is cleaned, we use a watershed algorithm to produce the boundary of the wing (Figure S2d). Watershed algorithms find catchment basins in an image, treating intensity as a height of the landscape. Since wings were detached from the body by breaking the hinge region with tweezers, each wing has a slightly different hinge due to experimental manipulation. Therefore, we use two morphologically identifiable landmarks, the humeral break and the alula notch (stars in Figure S2d). These landmarks are manually selected for each segmented wing image, and a script automatically defines the line between the two points as the proximal boundary. This defines a precise and repeatable way to delete the hinge. Area of each wing is measured as pixel number inside the wing boundary. Distribution of outbred group wing areas is shown in Figure S5g-j.

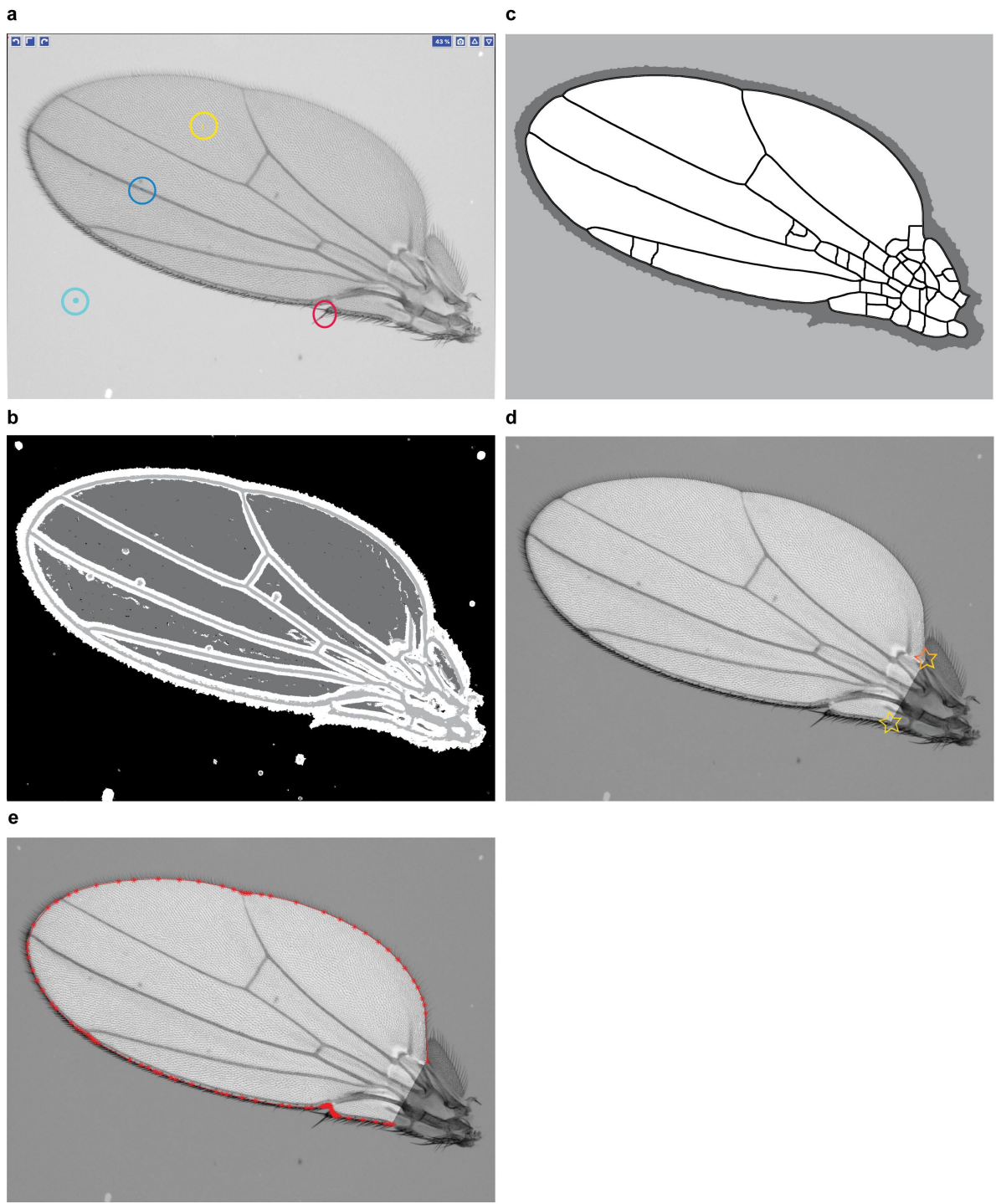

**Figure S2. Wing boundary identification. a,** An example of a training layer for Ilastik classification. The model has four layers: wing bulk (yellow), veins (blue), bristles (red), and background (cyan). **b,** Output of Ilastik. **c,** Cleaning the layers is required for boundary identification. Therefore, spurious veins appearing on the mask are not a concern. **d,** Example of a wing image after the wing boundary has been segmented and cleaned. The interior of the wing is unmasked, whereas pixels outside of the wing have been uniformly dimmed. This is done merely for demonstration purposes and is not part of the pipeline. Note there are two morphological landmarks marked with stars: the alula notch and humeral break. The complete boundary segmentation draws a line between them and excludes the hinge **e,** A polygon with vertices (red points) that we use to represent the wing boundary for conformal mapping.

**Conformal Mapping**

One feature of a conformal map is that it is possible to map any closed region onto another closed region if the shape is approximated as a complex polygon. For example, when we approximate the boundary of the *Drosophila* wing as a 100-point polygon, we can map the whole wing onto a unit disk. We distribute these points on the wing boundary with a density that is inversely proportional to the local curvature of the wing boundary (Figure S2e). Therefore we compute 2-dimensional curvature for every point of a segmented wing boundary using the `line-curvature2D` package[46]. Once we have curvature for every point, we compute and plot the cumulative curvature and distribute points equidistantly along the cumulative direction. The cumulative curve is monotonous; therefore, we can find the position of the vertices on the boundary. This approach minimizes the difference between polygonal representation and the real boundary for a given number of points.

Once a polygonal representation is made of a wing boundary, Schwarz-Christoffel mapping generates a map from the interior of the polygon onto a unit disc. The conformal map scales each local region homogeneously, but the scaling constant depends on the position. In the case of the *Drosophila* wing, which is approximately the shape of an ellipse, the bulk of the wing is minimally distorted, while regions far from the centre are more distorted. For instance, several peripheral pixels may be mapped onto one pixel of the unit disc (Figure S4k). Particular distortion depends on the shape of the wing, therefore if we would like to decouple shape from a pattern, this feature comes in handy.

We use the Schwarz-Christoffel Toolbox for Matlab[47] to map each segmented wing onto a unit disc that has a diameter of 200 pixels (Figure S4i). Schwarz-Christoffel conformal mapping requires only the boundary and centre-point (Figure S4b) of the shape. The shape's boundary becomes the circumference and the shape's centre-point becomes the origin. This is done individually for all of the segmented wings in an ensemble.

Note that the unit disc is the same area for all segmented wings. A wing boundary is a polygon in a complex space. Therefore, conformal mapping a segmented wing image onto the unit disc depends on the size of the segmented wing image. However, the target (the unit disc) has a fixed size, and so it may be the case that several pixels in the wing image may be mapped onto one pixel in the disc. Consequently, the intensity of the target pixel is an average of the wing image pixels that map to it. For the imaging resolution that was used, this happened most often for peripheral pixels (Figure S4k).

The Riemann mapping theorem implies that there is a biholomorphic (i.e. a bijective holomorphic mapping whose inverse is also holomorphic) mapping *f* from a non-empty simply connected open subset, $U$, of the complex plane $\mathbb{C}$ onto the open unit disk, $D = {z \in \mathbb{C}: |z| < 1}$.

The Schwarz–Christoffel formula is a recipe for a conformal map *f* from the upper half-plane (the canonical domain) $z \in \mathbb{C}: Im z > 0$ to the interior of a polygon (the physical domain). The function *f* maps the real axis to the edges of the polygon. If the polygon has interior angles $\alpha_1, \alpha_2, \alpha_3, \ldots$ , then this mapping is given by

$$f(z) = f(z_0) + C \int_{z_0}^{z} \prod_{j=1}^{j=1}(w - z_j)^{(\alpha_j/\pi)-1} dw$$

where $C$ is a constant, $\alpha_j$ are interior angles at the vertices, and $z_j$ are pre-images of the vertices, i.e. dots on the real axis. They obey inequality $z_1 < z_2 < \ldots < z_n = \infty$. For wing data, the unit disc contains 31,416 pixels, which is smaller in number than the ~100,000 pixels in each segmented wing image. The reduction was achieved by reducing the weight of the peripheral pixels and as a result, the shape of the wing has a lower weight.

**Conformal Map Alignment**

Since wing shapes and orientations are variable, the mapped wings must be globally aligned with one another. To align the mapped images, we use the remaining freedom, which is a rotation of the disc, plus we use the movement of the origin (Figure S4j). This freedom is an automorphism of the unit disk,

$$g(z) = e^{i\theta} \frac{z - \alpha}{1 - \overline{\alpha}z},$$

where $\theta$ is a rotation angle and $\alpha = g^{-1}(0)$ is a movement of the origin. We perform an independent parameter sweep of angle and displacement for one mapped wing image to obtain maximal cross-correlation between it and a reference wing disc. This operation is repeated for all of the other mapped wings, aligning each one to the same reference wing disc. We wrote a script that uses the Geometric Transformation Toolbox in Matlab to perform this procedure.

Since we are comparing variation within a group as well as variation between groups, we randomly choose one mapped wing to be the reference for an entire experiment. There is only one reference wing disc for all groups within the experiment varying diet and temperature (`G_25C_high_F_139_R`). There is only one reference wing for all groups within the experimental data comprising the genetic mutants (`G_samw_lei4X_F_110_L`).

Left and right wings have a different orientation. We use the function `fliplr` from Matlab to flip all images of left wings prior to conformal mapping. Once flipped they are treated in the same way as right wings, but we store information about their chirality. To determine if left and right wings are different in shape or pattern, we computed the Kullback-Leibler divergence (for details see SI.4.14) between left and right ensembles for the outbred group and found no difference. We also plot PC1 vs PC2 for the outbred group and use colour labels for chirality (Figure S5d). We find no correlation between the score for PC1/PC2 and chirality. These analyses indicate that there is no statistical difference between the left and right wings. All subsequent analysis aggregates both left and right wing data.

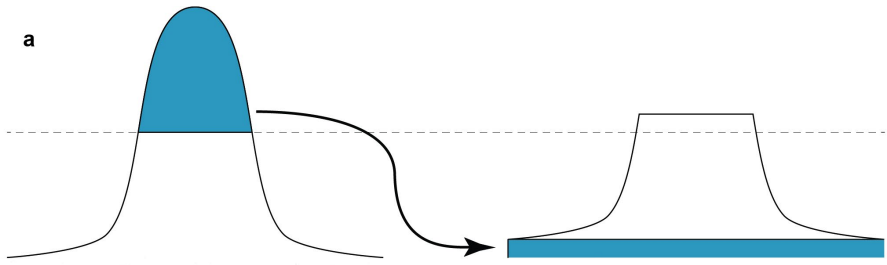

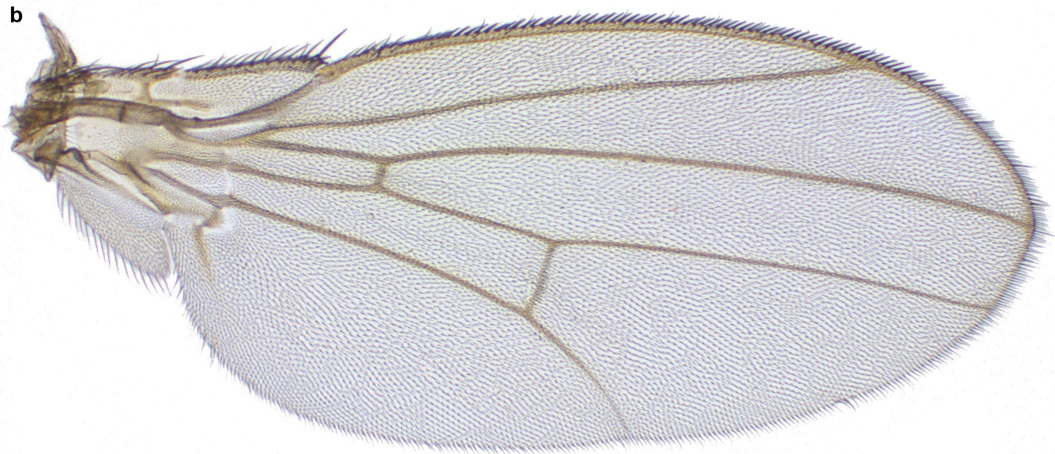

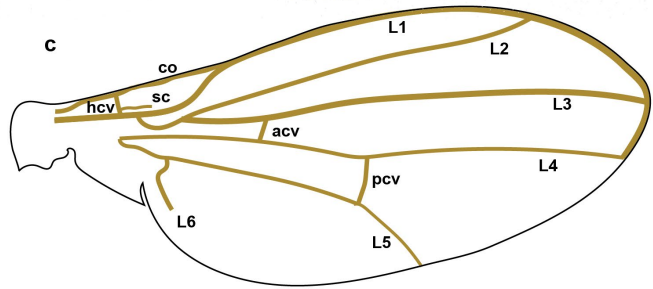

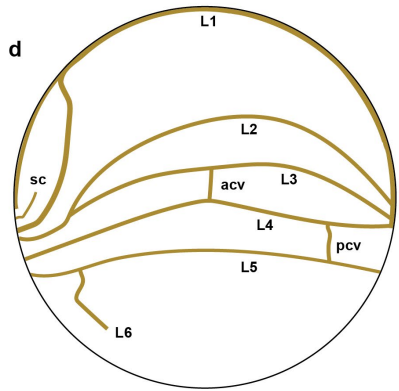

**Figure S3. Structure and form of the *Drosophila* wing. a,** Contrast-limited adaptive histogram equalization. The part of the histogram that exceeds the clip limit redistributes equally among all histogram bins. In contrast to ordinary histogram equalization, CLAHE is applied locally to patches of the image (tiles), and therefore uses several local histograms to redistribute the lightness values of the image. Thus, it is suitable for edge enchantment and improvement of the local contrast in every tile. **b,** A representative example of a male right wing from the outbred wildtype population that was imaged. **c,** Schematic of the wing structure. The longitudinal veins include: L1, L2, L3, L4, L5, and L6. The crossveins include the anterior (acv), humeral (hcv), and posterior (pcv) crossveins. Other veins include the costa (co) and subcosta (sc) veins. **d,** Schematized conformal map of wing in panel c showing the approximate locations and paths of the various veins.

**Radon Transformation**

The *Drosophila* wing is comprised of a sparse pattern of dark veins and lighter inter-vein compartments. We anticipate that much of the variation between wings will be detected as differences in vein thickness, position, and their angle relative to the body axes. A Radon transformation of a density function *f(x,y)* produces a function $Rf(L) = \int_L f(x)|dx|$ defined on a space of all lines *L*. For 2D images, a Radon transformation generates sinograms with coordinates of angle and *R* (Figure S6a,f-k). Since veins resemble lines, comparing two lines that differ by an angle is more sensitive to Radon-transformed data (Figure S6b-e). For this reason, we carry out all further analysis of the data with Radon-transformed images of aligned wing discs, obtained using the radon function in Matlab. However, when we need to present images of a wing, we use an inverse transformation `iradon` in Matlab.

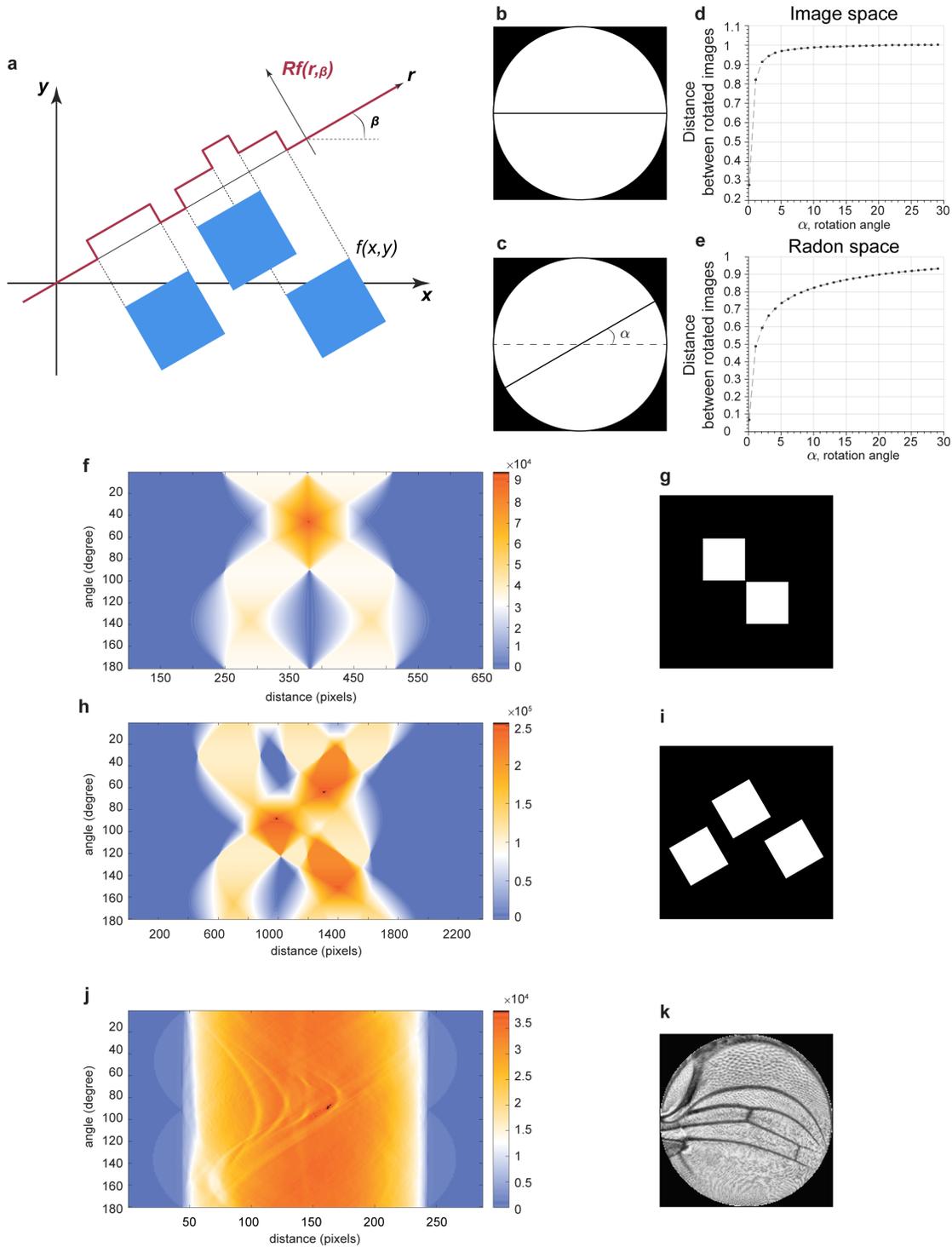

**Figure S6. Radon transformation. a,** A Radon transform maps $f$ on the $(x, y)$ domain to $Rf(r, β)$ on the $(r, β)$ domain. The function $f$ is equal to 255 in the blue squares and zero otherwise. This simulates an 8-bit image containing only the minimum and maximum intensity values. A transformation of $f$ onto a line with an angle of $β$ generates values of $Rf$ that vary as a function of $r$, the position along the line. Note that $Rf$ values are not binary, unlike the original image. The range and resolution of $Rf$ values is much greater. **b-e,** How a Radon transform makes line variation easier to detect. The reference image is a simple horizontal black line on the white disk (**b**), while the test image used for comparison is the same line but rotated by angle $α$ (**c**). The angle is systematically rotated in the test image, and the distance between test and reference images is measured and plotted. Distance is normalized to the distance between the reference image and the test image rotated by 90°. This analysis was performed in image space (**d**) and Radon space (**e**). Note how distance in image space is switch-like in dependence on angle as opposed to the dependence in Radon space. **f-i,** Example images binary images before Radon transformation (**g, i**). Radon transforms of the images showing $Rf$ values as heat maps on the plots of distance $r$ in pixel units and angle in degrees (**f, h**). **j-k,** A contrast-enhanced (CLAHE) wing image conformally mapped to a disc (**k**). $Rf$ values are shown as a heat map on the plot of distance $r$ in pixel units and angle in degrees (**j**).

**Analysis of Pattern Variation Within a Group**

Once all images are aligned to a reference on the disc, the remaining variability reflects biological variation. We use 8-bit grayscale images, which means that each pixel has intensity ranging from 0 to 255. Therefore, a wing image with N pixels can be represented as a single vector in discrete *N−*dimensional space, or to be more precise it is a vector in an N-dimensional cube ($[0,255]^N$). For a group with M number of wings, the group can be represented as a set of M vectors in *N−*dimensional space. Therefore, we can use all the tools from linear algebra.

We can analyse variability within a group since each group of individuals is subject to natural variation. We analyse this variation using PCA, which finds the directions of greatest variation within the group. Each principal component or PC is a unit vector in the same N-dimensional space as the wing discs comprising a group. We use the `pca` function from the Dimensionality Reduction and Feature Extraction toolbox for Matlab for all PCA-related computations.

We can also analyse variability within a group using Entropy. Each pixel at a fixed position may have different intensities for wings in an ensemble. If the pixel belongs to a region that has almost no variability in intensity, then it will have small entropy, but in the regions where there is pattern variability, the entropy will be larger. As an example, we can consider vein motion, which means that the position of the vein along either AP or PD axes varies between wings. If a pixel at a fixed position has identical intensity values for all aligned images, then that pixel has zero entropy. If a fixed-position pixel is within a wing vein for some but not all aligned images, then it might have larger entropy. In the intervein regions of wing images, the intensity of fixed-position pixels often varies slightly, so their entropy will be small but nonzero. We use the Kozachenko-Leonenko[31] entropy estimator for entropy

$$\widehat{H}(k) = log(c_d) + \log(N-1) - \psi(k) + \frac{d}{N}\sum_{i=1}^{N} log\, \rho_k(i),$$

where *d* is the dimension of x and $c_d$ is the volume of the *d*-dimensional unit ball $c_d = \pi^{d/2}/\Gamma(d/2 + 1)$ for Euclidean norm, $\rho_k(i)$ is the distance of the *i*-th sample to its *k*-th nearest neighbour, and ψ(x) is digamma function.

**Analysis of Pattern Variation Between Groups: Entropy-like Approach**

When there are two or more groups and we would like to compute the variability between them, we can estimate the Kullback- Leibler divergence, $D_{KL}$, or relative entropy. Let's consider two distributions *P* = *P*(*X*) and *Q* = *Q*(*Y*), the former one is interpreted as a real distribution (postulated prior probability) and the latter one is an assumed distribution that is measured at the experiment. The value of this functional is equal to the amount of information that is not accounted if *Q* be taken instead of *P*,

$$D_{KL}(P \parallel Q) = \sum_{i=1}^{n} p_i \log\frac{p_i}{q_i}.$$

We use the 2−nearest neighbour non-parametric Kozachenko-Leonenko estimator [48] for the Kullback-Leibler divergence,

$$D_{KL}(P \parallel Q) = \frac{d}{N}\sum_{i=1}^{N} \log\frac{\nu_k(i)}{\rho_k(i)} + \log\frac{M}{N-1},$$

where $\rho_k(i)$ is the distance from $x_i$ to its $k$-th nearest neighbour in $\{X_j\}_{j\neq i}$, and $\nu_k(i)$ is the distance from $x_i$ to its $k$-th nearest neighbour in $\{Y_j\}$.

**Analysis of Pattern Variation Between Groups: Centroid-based Analysis**

Another method to compare two or more groups is by centroid-based analysis. For one group containing Z wings, the group can be represented as a set of Z vectors in $N-$dimensional space. The centre of mass or centroid of the set can then be computed. The centroid can be thought of as the Mean Wing of the group (Figure S7a-b). Since we have a large number of wings for each group ($Z > 100$), the Mean Wing is a statistically robust representation of the group (also we used bootstrap to check robustness, see next section). We perform all manipulation for radon images of the wings and transform everything back with its inverse transformation.

We compute the Mean Wing for each group of wings, and then we can calculate the pairwise difference between Mean Wings of different groups. A pairwise difference is a vector that shows how one Mean Wing transforms into another Mean Wing. For example, when we say that there is a direction from Mean Wing *A* to Mean Wing *B* we mean that there is a vector, Δ, that generates Mean Wing *B* when added pixel-wise to Mean Wing *A*: *B* = Δ + *A*. If we normalize this vector, n = Δ/|Δ|, we get a unit vector that points from the reference group to another group (i.e., wildtype group to genetic mutant group). We can compare the direction of a unit vector with the direction of any PC vector from PCA analysis of a given group. This is done by computing a scalar product between the unit vector and the PC vector:

$$\cos\theta = (\boldsymbol{n}, \boldsymbol{PC}).$$

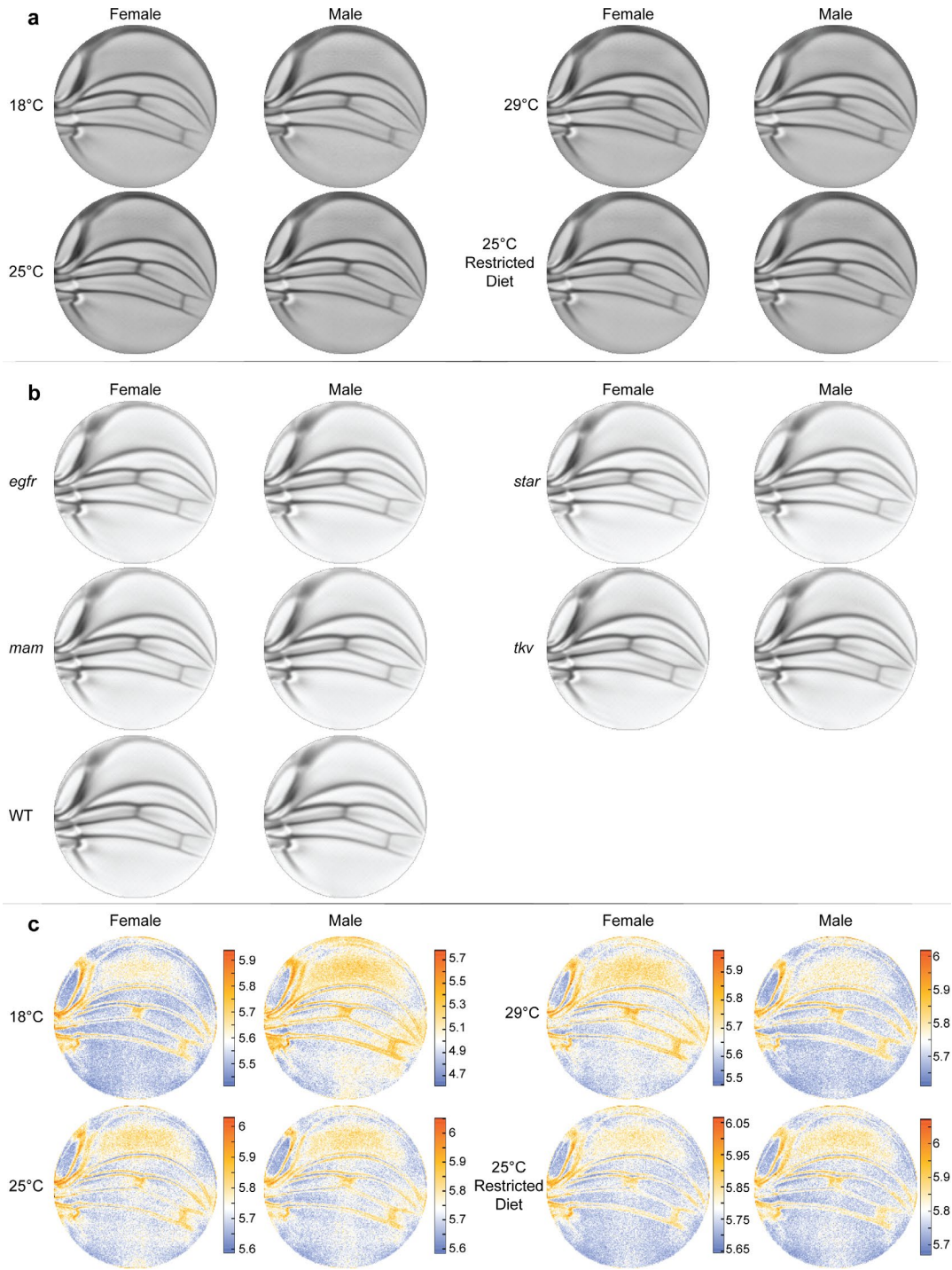

**Figure S7. Mean wing conformal maps from animals raised under different conditions or with different genotypes. a,** Mean wings from ensembles of outbred wildtype male and female flies raised under the diet and temperature conditions, as indicated. Differences between mean wings are subtle and require quantitative analysis to discern differences. **b,** Mean wings from the ensembles of wildtype (WT) and heterozygous mutants, as indicated. Note how qualitatively similar the mean wings are to one another by eye. **c,** Per-pixel variation within the ensembles shown in panel a are displayed as entropy plots. The Kozachenko-Leonenko 2nd-nearest neighbour estimator was used.

**Bootstrapping and Shuffling**

We have to check that the findings of the analysis are statistically robust. This requires that the populations are well-sampled, i.e. we have enough images so that the in-sample variance of the mean wing is small, and as a result that measured direction between populations is stable and close to the ideal direction that would be achieved if we had an infinite number of wings. For this approach, we use the bootstrap method that requires that we randomly draw subsets of wings that have a size of 50% of the ensemble. We repeat this procedure for each ensemble individually, so we do not mix wings that correspond to different populations. For pairs of such samples, we compute mean wings and the directional vector between them. We found that the numerical value of the $\cos\theta$ is stable, and the difference between sampled in this way $\cos\theta$ and value that is computed for whole populations is extremely low (few per cent of the corresponding values), suggesting that it is well estimated.

The second important check is to confirm that the trends we observe are biological in origin and not artefacts of our analysis. We perform a shuffle test. For this check, we construct new populations from images drawn randomly from the entire ensemble, which means that we mix wings from different populations (difference environmental and genetic conditions). We find that for these shuffled populations, a variation of the $\cos\theta$ is relatively large, and the mean value is only a few per cent of the unshuffled data. This suggests that the observed phenomenon depends on biological labels of the wings rather than artefacts introduced by the method and analysis itself.

**1D-Structure of the Observed Data**

While the data is embedded in high-dimensional space, it does not imply that the structure of the data is high dimensional as well. The simplest structure of the data that we may have is a one-dimensional cylinder, where the data is distributed along a line with noise in directions orthogonal to it. The simple structure of the data would hint at a simplifying principle that could be modelled and more deeply understood.

The angle between the **n** and PC vectors resides in an N-dimensional space. For a one-dimensional cylinder, one can compute a dependence of $\cos\theta$ as a function of $R$, the distance between points. If the distance between the two Mean Wings is $R$, then

$$\cos\theta = \sqrt{1 - \left(\frac{\sigma}{R}\right)^2} \Rightarrow \frac{\sigma}{R} = \sqrt{1 - \cos^2\theta}.$$

Thus, we get that $\sqrt{1 - \cos^2\theta} \xrightarrow{R \to \infty} 0$ as $\sigma/R$, which suggests the presence of a long direction in the data that dominates all other directions of variation. It further suggests that were $\cos\theta \sim R^{-1}$, then the noise in the directions orthogonal to the cylinder is constant along it. The simplest explanation for a different exponent would be that noise level increases with distance, and the structure is a cone rather than a cylinder. We experimentally measure the value of angles $\theta$ for pairs of bootstrapped centroids, computing $\sqrt{1 - \cos^2\theta}$ and we found that it qualitatively displays a decay as a power of $R$: $\sqrt{1 - \cos^2\theta} \xrightarrow{R \to \infty} 0$ as $\sigma/R^\alpha$. Values of α from fitting to an experimental data are reported in Figure S8.

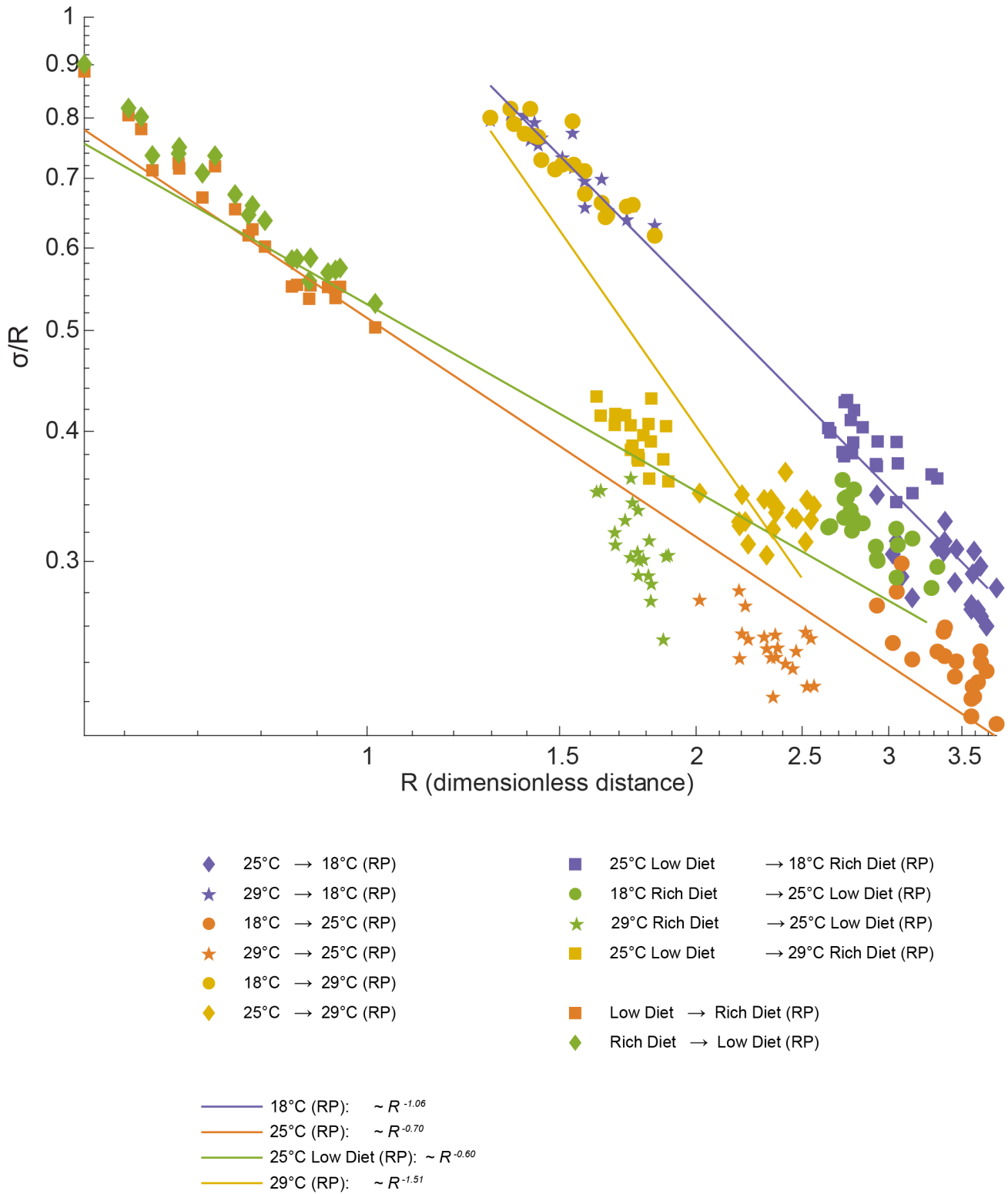

**Figure S8 Geometric analysis of dominant mode.** We observe the expected increase in alignment between directions of variation and directions connecting distinct populations. We fitted powers laws to the experimental value of the trigonometric quantity $\sqrt{1 - cos^2\,\theta}$ . One dimensional structure may be cylinder (if $\alpha = 1$) or a cone-like structure if $\alpha \neq 1$. An exact cylinder case means that "noise" is constant along the axes, while other power means that "noise" changes with distance in this space.